\documentclass[default]{sn-jnl}

\usepackage{graphicx}%
\usepackage{multirow}%
\usepackage{amsmath,amssymb,amsfonts}%
\usepackage{amsthm}%
\usepackage{mathrsfs}%
\usepackage[title]{appendix}%
\usepackage{xcolor}%
\usepackage{textcomp}%
\usepackage{manyfoot}%
\usepackage{booktabs}%
\usepackage{algorithm}%
\usepackage{algorithmicx}%
\usepackage{algpseudocode}%
\usepackage{listings}%
\usepackage{array}
\usepackage{adjustbox}
\usepackage{rotating}
\usepackage{pdfpages}
\usepackage{tabularx}
\usepackage{pdflscape}
\usepackage{makecell}
\usepackage{lineno}
\usepackage{todonotes}
\usepackage{pdfpages}

\theoremstyle{thmstyleone}%
\theoremstyle{thmstyletwo}%

\theoremstyle{thmstylethree}%

\newcolumntype{P}[1]{>{\raggedright\arraybackslash}p{#1}}

\begin{document}

\title[Article Title]{Enhancing Accessibility of Rural Populations \\ through Vehicle-based Services}


\author*[1]{\fnm{Clemens} \sur{Pizzinini}}\email{clemens.pizzinini@tum.de}

\author[1]{\fnm{Nils} \sur{Justen}}\email{nils.justen@tum.de}

\author[1]{\fnm{David} \sur{Ziegler}}\email{david.ziegler@tum.de}

\author[1]{\fnm{Markus} \sur{Lienkamp}}\email{lienkamp@tum.de}

\affil*[1]{\orgdiv{Institute of Automotive Technology}, \orgname{Technical University of Munich}, \orgaddress{\street{Boltzmannstr. 15}, \city{Garching b. München}, \postcode{85748}, \state{Bavaria}, \country{Germany}}}


\abstract{Improving access to essential public services like healthcare and education is crucial for human development, particularly in rural Sub-Saharan Africa. However, limited reliable transportation and sparse public facilities present significant challenges. Mobile facilities like mobile clinics offer a cost-effective solution to enhance spatial accessibility for the rural population.
Public authorities require detailed demand distribution data to allocate resources efficiently and maximize the impact of mobile facilities. This includes determining optimal vehicle service stop locations and estimating operational costs.
Our integrated approach utilizes GIS data and an accessibility scaling factor to assess spatial accessibility for rural populations. We tailor demand structures to account for remote and underserved populations. To reduce average travel distances to 5 km, we apply a clustering algorithm and optimize vehicle service stop locations.
In a case study in rural Ethiopia, focusing on four key public services, our analysis demonstrates that mobile facilities can address 39-62\% of unmet demand, even in areas with widely dispersed populations.
This approach aids decision-makers, including fleet operators, policymakers, and public authorities in Sub-Saharan Africa, during project evaluation and planning for mobile facilities. By enhancing spatial accessibility and optimizing resource allocation, our methodology contributes to the effective delivery of essential public services to underserved populations.}

\keywords{Rural Accessibility, Vehicle-based Services, Location-Allocation, Mobile Facilities}



\maketitle

\section{Introduction}\label{sec:introduction}
\begin{figure}[htb]
\includegraphics[width=\textwidth]{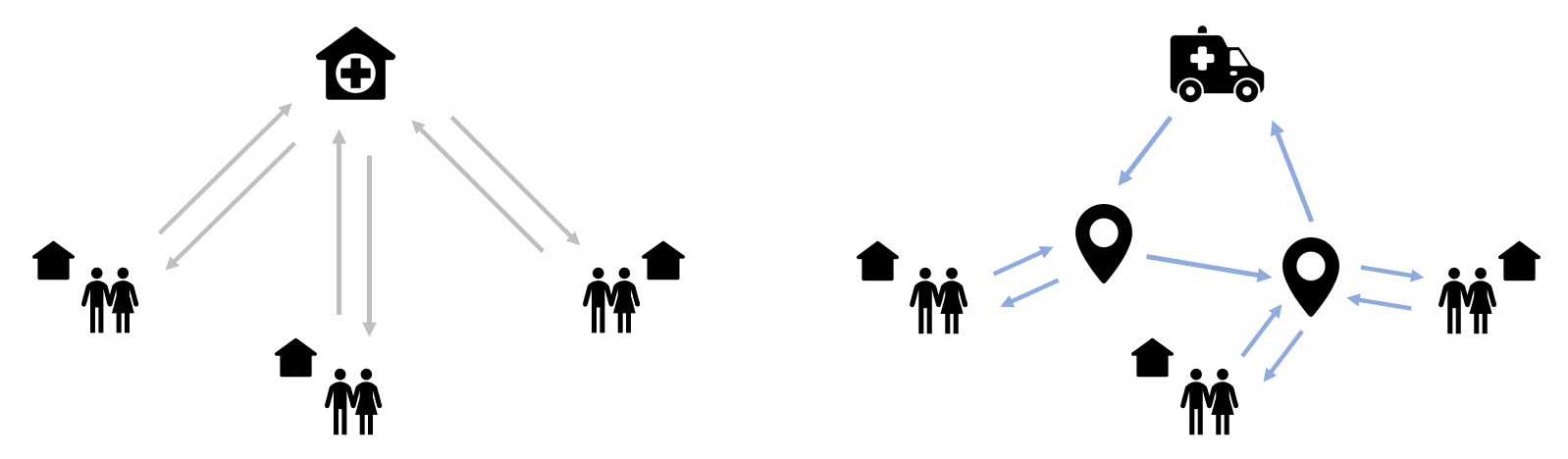}
\centering
\caption{Basic principle of a vehicle-based service (VbS) system}
\label{vbs_prinzip}
\end{figure}

Mobile facilities play a crucial role in enhancing access to a variety of public services that are integral to leading healthy, meaningful, and fulfilling lives \cite{Doerner2007MulticriteriaCountry, Bodenheimer1969MobileProblem, Raghavan2019TheProblem}. The concept of vehicle-based services (VbS) combines this variety of services that can be delivered utilizing vehicles (Figure \ref{vbs_prinzip}.)\cite{Pizzinini2022FromServices}. Prominent international organizations have previously advocated for the effective deployment of VbS to provide essential public services in rural areas, with a particular focus on vulnerable populations in Sub-Saharan Africa (SSA). Numerous studies have highlighted successful VbS projects encompassing healthcare, vaccinations, and dental hygiene \cite{Oladeji2021ExploringEthiopia, Sharma2011AnIndia, Fox-Rushby1996CostsGambia, Rudolph1992AAfrica, Swaddiwudhipong1999AWomen}. While it may be reasonable to question whether VbS can offer the full spectrum of services comparable to large hospitals or shopping centers, many of these interventions are relatively straightforward yet hold the promise of significant enhancements over the existing status quo \cite{Starkey2007AServices}.

In the process of project evaluation, public authorities face the challenge of comparing investment and operating costs against effectiveness when considering both stationary and mobile facilities \cite{Thacker2019InfrastructureDevelopment}. This necessitates the determination of potential service locations. While location-allocation models for stationary infrastructure have been extensively explored in the literature \cite{Pu2020ImprovingCongo, Gould1966ApproachServices, Chung1986RecentModel}, the selection of optimal vehicle service stop locations for VbS has received limited attention in research until the present day. Identifying these locations is critical for making strategic, tactical, and operational decisions regarding fleet size, vehicle types, and vehicle routing. This information is essential for conducting a-priori cost-benefit analyses \cite{Raghavan2019TheProblem, Bodenheimer1969MobileProblem, Halper2015LocalProblem}. 

This paper contributes to the body of literature in two ways. First, it addresses a gap in existing research on location-allocation models, where the assumption of fixed candidate locations for both stationary and mobile facilities is prevalent \cite{Pu2020ImprovingCongo, Raghavan2019TheProblem, Halper2015LocalProblem, Doerner2007MulticriteriaCountry}. Such an assumption implies the existence of explicit regulations, well-defined property rights, and known infrastructure restrictions, which are often not the case in rural SSA \cite{Pu2020ImprovingCongo, Gould1966ApproachServices}. This holds explicitly true for the location-allocation of hospitals. Existing healthcare-focused approaches take zoning regulations, local ordinances and infrastructure access as their initial input \cite{Weiss2020GlobalFacilities, Yao2013AAfrica, Pu2020ImprovingCongo}. Applying these approaches for other public services would hinder an optimal placement in rural SSA regions. To address this, we propose a demand clustering algorithm based on population distribution data only. In this way, we can establish vehicle service stops in undeveloped areas.

Second, only considering the inhabited regions, many SSA countries have relatively evenly distributed populations \cite{TheInternationalFundforAgriculturalDevelopmentIFAD2023GlobalPopulation} but existing approaches tend to focus on locating facilities close to population centers to maximize cost-effectiveness \cite{Rahman2000UseNations, Farahani2012CoveringReview}. However, this utilitarian perspective may overlook the most vulnerable individuals who are often scattered across vast areas. Our algorithm incorporates an Egalitarianism Principle, aiming to ensure equitable outcomes for these neglected populations \cite{Feitosa2021SpatialOpportunities}.

In the following sections, we outline a step-by-step approach to selecting and prioritizing vehicle service stop locations for VbS, including a comprehensive analysis of healthcare, education, energy, and water access in our target region in Ethiopia. This paper is organized as follows: the second section briefly discusses general spatial accessibility measures and introduces existing efforts to locate public facilities. Next, we present our proposed methodology followed by an outline of our study in Bekoji. The results section presents the analysis of vehicle service stops for the case study, including a sensitivity analysis and robustness check. We conclude with a critical discussion of the method's applicability and potential limitations.

\section{Related work}\label{demand_to_supply}

\subsection{Measuring spatial access}\label{sec:measure_access}
Various measures can be deployed to understand a population’s access level. The more parameters associated with a given model, the more difficult it is to implement, and the less transferable it is across regions with potentially less available data. The following excerpt of popular models briefly explains the methodology and analysis their shortcomings for our application.

\subsubsection{Gravity model}
Gravity models assume that the spatial accessibility of a population to service locations declines with increased distance \cite{Wan2012AServices}. These models have been used in various accessibility analyses and assess the spatial interaction between any population point and all service points by discounting the potential of a given service with increased distance or increased travel impedance \cite{Gharani2015AnStates, Wan2012AServices, Guagliardo2004SpatialChallenges}. The accessibility of a given service is assumed to improve if the quantity of service locations increases, the service capacity at any one location is increased, the travel distance to a facility is reduced, or the corresponding travel friction decreases \cite{Gharani2015AnStates}. 

Yao et al. \cite{Yao2013AAfrica} use a gravity-based model to assessed access to sexual and reproductive health care for women in rural SSA. In their two-step process, a measure for the service quality of a health facility is calculated and incorporated into a gravity-based model to understand the likelihood that inhabitants would travel to a respective health facility.

Nelson et al. \cite{Nelson2008Estimated2000} apply a cost-distance algorithm that accounts for landscape characteristics and transport infrastructure to compute travel times. This allows a comparison of different modes of transport and shortest-path considerations. Based on this procedure, the Global Accessibility Map computes a population gradient around large cities of 50,000 or more people. Following this approach, Weiss et al. \cite{Weiss2020GlobalFacilities} derive travel times to the nearest hospitals. They utilized data from Google Maps combined with the Global Accessibility Map and aggregated data on hospital locations. The authors could then depict the proportional distribution of the population relative to healthcare facilities. 

There are three drawbacks to the gravity model for our objective. Because it is necessary to discretely model population demand as points, often taken to be census tract centroids, this method suffers from edge effects, whereby it fails to account for the border crossing to seek service in a neighboring postal code, for instance. This becomes especially problematic for small geographic regions \cite{Guagliardo2004SpatialChallenges}. In rural Ethiopia, many of the administrative districts are over 30 km across. Even if data was widely available at a district level, and aggregated at its respective centroid, this would likely still be an inaccurate representation of regional demand and lead to a sub-optimal distribution of vehicle stop locations. Secondly, the assumption of fixed supply ratios does not incorporate variations in accessibility within an assumed access area \cite{Guagliardo2004SpatialChallenges}. As a result, gravity model results can vary widely and may be unintuitive to interpret \cite{Luo2009AnPhysicians}. Furthermore, calculating the friction coefficient is often tied to assumptions that harbor inherent uncertainties \cite{Luo2009AnPhysicians, Hu2013AssessingCounty}. Lastly, because the gravity model does not account for overlapping service area catchments, there is a tendency to overestimate the demand for service sites, an effect that becomes greater when more sites are present in a given region \cite{Wan2012AServices}.

\subsubsection{2-step floating catchment area approach}
The 2SFCA approach is, in essence, a service-to-population ratio in the form of floating catchment areas that are permitted to overlap \cite{Wan2012AServices}. This results in a more realistic utilization of behavior modeling \cite{McGrail2012SpatialImprovements}. The approach has been used to model various spatial access problems, focusing on healthcare applications \cite{Wan2012AServices}. In the first step, an initial service-to-population ratio centered at the respective service locations is computed \cite{Gharani2015AnStates}. Then, for every demand location's catchment, these rations are summed to derive a measure of accessibility with a larger value indicating greater access \cite{Gharani2015AnStates, Luo2009AnPhysicians}. 

Again, three limitations can be pinpointed here. First, the catchment areas are defined in terms of maximum distance or travel time \cite{McGrail2012SpatialImprovements}. Under this approach, all individuals within a catchment area are seen to have equal access to a service location, while all locations outside of the catchment count as completely inaccessible \cite{McGrail2012SpatialImprovements, Luo2009AnPhysicians}. This binary classification likely does not represent impedance behavior and becomes especially questionable for large catchments \cite{Hu2013AssessingCounty}. Second, regardless of distribution, populations are assumed to reside at a single aggregated point, often the centroid of an administrative tract \cite{McGrail2012SpatialImprovements}. While this assumption may hold in cases where data is readily available, and census tracts are small, the assumption becomes questionable in rural settings with a more uniform and dispersed population distribution often found in less urbanized regions in SSA. Lastly, the model assumes that catchment sizes, whether computed in time or distance, are identical across services and populations, which may introduce errors based on observed behavioral patterns \cite{McGrail2012SpatialImprovements}. It is to be expected, for instance, that individuals are likely willing to travel further for some services - such as cancer treatment - than for others - a dentist appointment.

\subsubsection{Enhanced 2-step floating catchment area approach}
The enhanced 2SFCA model builds on the original by adding distance decay weights to the previously binary accessibility classification, addressing the assumption of non-diminishing access within a catchment area \cite{Luo2009AnPhysicians, Wan2012AServices}. As a result, access to service now decreases with increased travel time. By adding decay weights, the approach reveals more finely-grained accessibility classifications and mitigates the previous concern of overrated accessibility for populations residing at catchment edges, which led to demand overestimation under the 2SFCA model \cite{Luo2009AnPhysicians}.

Despite the marginal improvement over the 2SFCA, many original shortcomings remain. While the travel time catchment subdivisions are accurate to a point, the model continues to assume that all those individuals within a subdivision have equal access; in other words, no individuals in one region venture to another to seek out a service \cite{Luo2009AnPhysicians, McGrail2012SpatialImprovements}. In addition, the impedance coefficient selection or calculation makes a series of potentially problematic assumptions about the utilization behavior of different populations, introducing further uncertainty into the model \cite{Hu2013AssessingCounty}. To date, there is limited literary consensus around the appropriate method for selecting beta, with some studies advocating calculation according to a predetermined mathematical expression, and others positioning themselves in favor of selecting fixed values \cite{Hu2013AssessingCounty, McGrail2012SpatialImprovements}. Lastly, as with the 2SFCA approach, the enhanced version continues to assume that populations are aggregated at a single point, which may not be accurate in rural settings in SSA \cite{McGrail2012SpatialImprovements}.

\subsubsection{3-step floating catchment area approach}
The 3-Step Floating Catchment Area approach adjusts the original model further by attempting to address the fact that a population’s demand for a given service is influenced at least in part by the availability of other nearby service centers \cite{Wan2012AServices}. This attempts to overcome, at least in part, the demand overestimation problem associated with the 2SFCA approach \cite{Wan2012AServices}. To accomplish this, the model effectively implements a competition scheme, adding a selection weight to the methodology, computed by dividing each weight by the sum of all weights \cite{Wan2012AServices}. The approach remains otherwise identical. In model evaluations, the 3SFCA approach was indeed found to have a moderating effect on service demand projections, leading to a more accurate prediction of respective shortage areas \cite{Wan2012AServices}. The remaining weaknesses of both the enhanced and the original 2SFCA models remain unaddressed.

\subsection{Location-allocation models}\label{sec:localtion_allocation_models}
Location-allocation models aim to find optimal sites for supply facilities based on spatial access measures. Until today, a variety of models have been suggested in the literature. In the following, three of the most relevant methods are briefly explained. P-median problems, location set covering problems, and maximum covering location problems. All of these generally fall under the category of node-based approaches.

\subsubsection{P-median model}
The P-median model is popular among analysts and aims to minimize the weighted distance between a requesting node and the nearest facility \cite{Baouche2014EfficientNetwork}. This is accomplished by locating a given number of facilities over a geographic area and allocating demand nodes to these facilities to minimize the total consumer distance or time traveled \cite{Deb2018ReviewVehicles, Rahman2000UseNations}. The model requires the explicit modeling of demand nodes via the selection of candidate location sites. Like the majority of node-based approaches, the assumption here is that users make only simple trips from their home to a respective utility; this exposes another inherent shortcoming of not being able to accurately represent more complex trip patterns \cite{Deb2018ReviewVehicles}. 

Its application in rural areas dates back to the 1960s and the allocation of hospitals in Guatemala \cite{Gould1966ApproachServices}. The solution space for an optimal supply site selection was limited to a discrete network or continuous space and is formulated as p-median (reduced total travel times between demand points and a new location) \cite{Gould1966ApproachServices}. 

\subsubsection{Location set covering model}
Location set covering models aim to find the lowest number of facilities such that every demand is covered by at least one facility within a pre-specified distance or travel time \cite{Deb2018ReviewVehicles, Baouche2014EfficientNetwork, Church1974TheProblem, Church2003TheModel, Chukwuma2018GISNigeria}. Despite the limited input data these models generally require, it remains necessary to define an explicit number of discrete demand nodes \cite{Church2003TheModel, Straitiff2010UsingPlane, Chukwuma2018GISNigeria}. Notably, there are few literary references to continuous coverage problems, with most research limited to discrete applications \cite{Farahani2012CoveringReview}. Consequently, a tradeoff exists between computation time and the scale of representation, which may be problematic for sparsely populated rural regions in SSA \cite{Straitiff2010UsingPlane, Farahani2012CoveringReview}. A comprehensive literature review of facility covering problems by Farahani and Asgari et al. \cite{Farahani2012CoveringReview} finds that the majority of these approaches continue to stipulate the definition or simplification of discrete demand data; this usually results in a summation of points, and inevitably leads to aggregation errors and loss of information.

\subsubsection{Maximal covering location model}
The Maximal Covering Location Problem seeks to maximize demand coverage by accounting for circumstances that may call for a given demand node to be covered by more than one facility. A scenario in which the nearest ambulance to an emergency is busy on another call, in which case the second-closest ambulance from a different fire station should still be able to reach the location within a specified amount of time illustrates the model \cite{Farahani2012CoveringReview}. The basic parameters, and drawbacks, remain unchanged to the previously introduced location set covering problem \cite{Baouche2014EfficientNetwork, Church2003TheModel}.

\subsection{Spatial justice considerations}
Spatial accessibility assessment comes with an ethical dimension. While private facilities are understandably driven by profit-maximizing objectives, public services are on their surface committed to a certain level of equity. Despite this assumption, investments of this type also tend to concentrate in central locations \cite{Feitosa2021SpatialOpportunities}. Given the strong correlation between spatial accessibility and a host of quality-of-life indicators, this concentration is essential. The first of these is the utilitarian approach. This approach treats all individuals as equals, disregarding personal and situational differences; the subsequent potential benefits of a given resource distribution are then evaluated as equal among all individuals, to maximize aggregate welfare \cite{Feitosa2021SpatialOpportunities}. In other words, utilitarian approaches aim to achieve the greatest possible benefit for the most significant number of people \cite{Feitosa2021SpatialOpportunities}. The Rawlsian principle looks to limit the total level of inequality caused by a specific spatial distribution of resources. According to Rawls, inequalities should be structured to confer the most benefit to the least well-off, thereby considering a given individual’s varying initial conditions based on accessibility \cite{Feitosa2021SpatialOpportunities}. In practice, this approach stresses the importance of focusing efforts to improve spatial accessibility on those whose initial position is worst. While the utilitarian approach would prefer to locate a service location closest to the most significant number of people, Rawl’s Difference Principle ensures the greatest benefits for those least well-off.

\subsection{Vehicle service stop allocation}
VbS is a novel concept that sparks interest from practitioners and researchers to enhance spatial accessibility to public services in rural regions \cite{Pizzinini2023DrivingAfrica, Wildman2013TechnicalEmergencies, Hill2014MobileReform.}. Whereas the quality of public service delivery with mobile facilities might be inferior (i.e., enclosed room available for medical treatment), the characteristics of vehicle service stops yield some advantages over stationary facilities:
\begin{itemize}
    \item \textbf{Setup time}: Mobile facilities substantially reduce setup time. This is particularly important during a time-sensitive emergency and disaster intervention \cite{Wildman2013TechnicalEmergencies}, but also enables irregular and on-demand service deliveries in rural communities \cite{Bodenheimer1969MobileProblem}. 
    \item \textbf{Relocation}: Vehicle service stops can be relocated continuously according to demand and/or supply patterns \cite{Alarcon-Gerbier2022ModularReview}. 
    \item \textbf{Location flexibility}: Mobile facilities for public service delivery required a high degree of autarchy (i.e., on-board energy and water supply)  \cite{Pizzinini2023DrivingAfrica, Pizzinini2022FromServices}. This renders VbS particularly flexible to operate in regions with no existing infrastructure. 
    \item \textbf{Versatility}: Vehicle service stops only represent a dedicated location for community members to access mobile facilities. Once located, several different services can be offered at the same location \cite{Pizzinini2023DrivingAfrica}.
\end{itemize}

All these characteristics relate to the overall increased operational flexibility that can be achieved with mobile facilities. Whether or not this flexibility is cost-effective for a particular intervention area must be assessed individually \cite{Pizzinini2023DrivingAfrica}. However, adopting existing allocation principles might cannibalize the outlined temporal and spatial flexibility of VbS. The introduced location-allocation models aim to find the optimal service location for fixed demand and supply patterns. Inherently to each of their optimization objectives (minimizing travel times \cite{Baouche2014EfficientNetwork}, minimizing number of facilities \cite{Deb2018ReviewVehicles, Baouche2014EfficientNetwork, Church1974TheProblem, Church2003TheModel, Chukwuma2018GISNigeria} maximize coverage \cite{Farahani2012CoveringReview}), these models are geared towards finding a permanent location for a brick-and-mortar supply site with a long-term planning horizon.

\subsection{Identified research gap}
Despite the need for a location-allocation model that aligns with the flexibility of vehicle service stops for VbS delivery, we find further gaps in the current literature on facility location-allocation. Firstly, the introduced models rely on given population thresholds for settlements or base their analysis on administrative divisions available before the project appraisal \cite{Baouche2014EfficientNetwork}. In particular, across SSA, such inquiries are often not available \cite{Nelson2008Estimated2000}. Further, demand and service candidate locations must be known before the appraisal \cite{Baouche2014EfficientNetwork}, limiting the general scalability of a potential method since data needs to be gathered. These approaches remain critically inflexible, too expensive for large-scale application, or incompatible with the sparse data available in rural settings. As a result, they do not lend themselves to practical application by policymakers and practitioners, where the ease of use and potential scalability for trans-regional analysis is essential \cite{OPPONG2012OBSTACLESAFRICA, Doerner2007MulticriteriaCountry, Rahman2000UseNations}. Further, no research proposes a methodology that can cope with data of different quality covering multiple public service interventions (i.e. water supply, electricity, education, etc.). Most of the published research applying the models focuses on accessibility to healthcare services were data quality is comparable between most SSA countries \cite{Weiss2020GlobalFacilities, Yao2013AAfrica, Pu2020ImprovingCongo}.  

Considering these shortcomings, deriving a flexible methodology that can work with incomplete data, does not require the specification of a limited number of population centers, and can draw on insights from a variety of geospatial databases has the potential to provide a valuable new perspective on rural accessibility problems, to highlight effective interventions. 

\section{Data and methods}

\subsection{Data sets and case study}
We utilize QGIS, an open-source GIS software for all the following operations. Further, we limit our analysis to publicly available datasets (Table \ref{table:dataset}) and always use the highest-resolution demand data available. This setup is replicable and inexpensive for public authorities and fleet operators in SSA. 

We apply our process to a test location in Ethiopia to showcase the methodology. The site is part of a research project with the German Agency for International Development (GIZ) \cite{TUM2022ACarFTM}. Two vehicles are stationed in Bekoji, in the state of Oromia, and operate daily within a range of 50 km along the existing road network. The VbS considered for this case study are healthcare, education, energy, and water service.

\begin{table}[htb]
\caption{Applied GIS datasets for the case study}
\setlength{\tabcolsep}{10pt} 
\renewcommand{\arraystretch}{1.5} 
\begin{tabular}{P{2.8cm}P{4.8cm}p{1.5cm}p{1cm}}
\hline
Dataset & Content & Resolution & Date \\ \hline
OpenStreetMap      & Base layer with road network and location data \cite{OSM2022OpenStreetMaps} & – & 2022 \\
WorldPop & Population and sociodemographic data \cite{Linard2012Population2010} \cite{Tatem2014MappingBirths} & 0.1x0.1km & 2020 \\
USAID DHS & Sociodemographic data (literacy rates, access to drinking water) \cite{USAID2022TheRepository} & 5x5km & 2016 \\
UN OCHA & Health facility data \cite{Ouma2018AccessAnalysis} & – & 2015 \\
NASA  & Night light data on electrification \cite{NASAEarthObservatory2022EarthMaps} & 3x3km & 2016 \\
Global Accessibility Map & Global data travel times to the nearest settlement over 50,000 inhabitants\cite{Nelson2008Estimated2000} & 1x1km & 2008 \\ \hline
\end{tabular}
\label{table:dataset}
\end{table}

\subsection{Service selection}
While our approach is tailored for SSA and applicable throughout all public services that can be rendered with vehicles \cite{Pizzinini2023DrivingAfrica}, we select four public services based on their importance for sustainable human development, their notorious undersupply in the study area, and their diverse requirements for being delivered by vehicles (i.e., water tank vs. mobile medical unit \cite{Pizzinini2023VehiclesService}). We combine service locations of four public services to leverage the full potential of VbS with vehicles capable of delivering multiple service types \cite{Pizzinini2023VehiclesService}. 

Across SSA 29\% of the population live more than two hours on foot from the nearest hospital \cite{Meara2015GlobalDevelopment, Broer2018GeospatialAfrica, Ouma2018AccessAnalysis}. Maternal mortality is a dramatic indicator of inequity in global health, and potential interventions for these services have some of the greatest expected marginal returns \cite{Tatem2014MappingBirths, Fox-Rushby1996CostsGambia, Rahman2000UseNations}. Across Ethiopia, the average individual lives 10 km from the nearest ante-natal care (ANC) facility, and in the regional state of Oromia where Bekoji is located, this increases to 14.5 km \cite{Tegegne2019AntenatalAnalysis}. Previous ANC services implemented in SSA offered in the form of mobile labs in the Gambia, have significantly improved maternal health \cite{Fox-Rushby1996CostsGambia}. We, therefore, define VbS 1: ANC as one ultrasound scan rendered by trained personnel, including an infection test.

In SSA, less than a third of people have grid access; in many rural areas, this drops to less than 15\% \cite{Louie2015RuralLeaders}. In Ethiopia, just 8\% of rural households have access compared to 93\% of urban households \cite{CentralStatisticalAgency2016EthiopiaFindings}. Vehicles can conceivably transport upwards of 140 fully charged portable battery kits, returning empty batteries to a central hub for overnight charging \cite{Louie2015RuralLeaders}. We define VbS 2: Energy as the delivery of one fully charged power kit for rental.

Across SSA countries, millions of children lack basic writing, reading, and numerical skills \cite{Oluwatobi2015MobilePoor, Hanjra2009PathwaysMarkets, Stifel2017MarketEthiopia}. In Ethiopia, only 42\% of students have access to textbooks in primary schools,and many facilities are unsuitable for year-round teaching \cite{Oluwatobi2015MobilePoor}. Mobile facilities offering education services \cite{Beyers2012MobileAfrica} and scholastic material delivery \cite{Evans2021EducationLearning, Niemand2021TheAfrica} have been recognized to have a positive impact on educational attainment. Therefore, our VbS 3: Education delivers one textbook and basic school supplies.

According to findings by the United Nations, the use of unsafe drinking water is four times more prevalent among rural communities, and in less accessible regions of Ethiopia, almost 40\% of the population lack access to clean water \cite{CentralStatisticalAgency2016EthiopiaFindings}. In many parts of SSA, water trucking has become a frequent intervention to ensure the supply of five liter of drinking water per person in drought-affected regions \cite{Wildman2013TechnicalEmergencies}. VbS 4: Water is the distribution of five liter bottled drinking water.

\subsection{Assumptions}
\label{assumptions}
Throughout this work, we follow two reasoned assumptions. First, we assume that people in the rural target area primarily travel by foot \cite{DosAnjosLuis2016GeographicMozambique, Fatih2013AccessibilityGIS}. Using a commonly referenced walking speed of 5 km/h, we thus arrive at a practical upper bound of total travel time \cite{Weiss2020GlobalFacilities, DosAnjosLuis2016GeographicMozambique, Broer2018GeospatialAfrica, Tobler1993ThreeModeling}. Simply assuming that communities have access to more advanced means of transportation, runs the risk of overestimating existing accessibility to these services and thereby unwittingly excluding vulnerable subgroups of the population \cite{Bodenheimer1969MobileProblem, Luo2009AnPhysicians, Stifel2017MarketEthiopia, Damsere-Derry2008AssessmentGhana}. 

Second, we utilize a distance-based cut-off measure of 5 km Euclidean distance between demand and supply points. It follows that populations within a 5 km radius, the so called catchment area, around a supply point are deemed to have access. This cut-off measure stems from studies by international agencies such as the World Health Organization \cite{WorldHealthOrganizationBackgroundSystems}. 
Notably, for this work, the wider body of literature does not unanimously agree on a set of criteria or cutoff points by which to differentiate accessible from inaccessible regions \cite{DosAnjosLuis2016GeographicMozambique, Tegegne2018TheMeta-analysis, Fatih2013AccessibilityGIS, Doerner2007MulticriteriaCountry}. 

Third, we utilize the Global Accessibility Map \cite{Nelson2008Estimated2000} to prioritize those populations that reside furthest from established population centers. Such centers are assumed to have an increased supply of public services  \cite{Nelson2008Estimated2000}. We therefore calculate a region-specific Accessibility Scaling Factor $A$. By dividing the friction value $d_{i}$ (ease at which humans can move through each pixel of the world’s surface) in a respective part of interest by the maximum impedance value $d_{max}$ (distance characterized by travel time) observed in this same area of interest. This factor can be calculated for each pixel $i$ in the region $I$. The value of the Accessibility Scaling Factor increases with an increasing friction value, increasing the weight of remote populations for the following calculations.

\begin{equation}
\begin{aligned}
\textrm{Accessibility Scaling Factor} \hspace{0.1cm} A_{i} = \frac{d_{i}}{d_{max}} \hspace{1cm} \forall i \in I
\end{aligned}
\label{accibilityscalingfactor}
\end{equation}
\vspace{0.2cm}

\subsection{Total demand quantification}
\label{demandquantification}

\begin{figure}[htb]
\includegraphics[width=\textwidth]{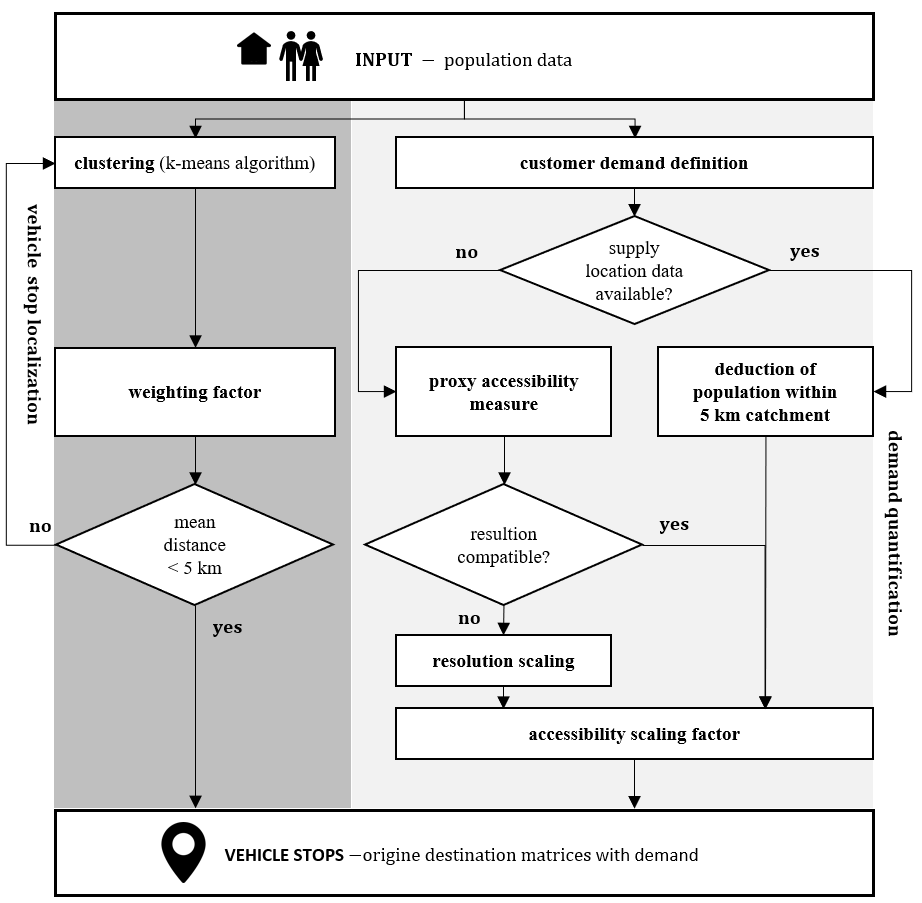}
\centering
\caption{Overview of the proposed methodology}
\label{process}
\end{figure}

Demand for each VbS is derived from WorldPop either directly or by arithmetic operations on additional datasets. For VbS 1: ANC, we assume all women of childbearing age (demographic) should receive three ultrasound scans during pregnancy (temporal). In this case, absolute demand can be directly derived from the WorldPop dataset. We define $p_{i}$ as the annual pregnancy counts at pixel $i$ and multiply it with the Accessibility Scaling Factor $A$.

\begin{equation}
\begin{aligned}
\textrm{VbS 1: ANC} = p_{i} * A_{i} \hspace{1cm} \forall i \in I
\end{aligned}
\label{ANC-calcuations}
\end{equation}
\vspace{0.2cm}

The number of potential customers for VbS 2: Energy is not available off-the-shelf and requires arithmetic operations on available GIS data. A combination of population data and NASA Night Light Data yields a proxy for the adult population without electricity for illumination at night. We define $p_{i}$ as the number of individuals and $a_{i}$ as night light intensity found at pixel $i$. We normalize the absolute values $a$ from the NASA data and scale it equally across the 100~m $*$ 100~m resolution population data $p$.

\begin{equation}
\begin{aligned}
\textrm{VbS 2: Energy} = p_{i} * \frac{a_{i}}{a_{max}} * A_{i} \hspace{1cm} \forall i \in I
\end{aligned}
\label{normalize_light}
\end{equation}
\vspace{0.2cm}

To determine the need for VbS 3: Education, we follow a similar approximation. The USAID data on gender-specific literacy rates, expressed as a percentage, is averaged to define an overall literacy rate $L$ in the region. After deriving the percentage of non-literate individuals from this distribution, the quantity is multiplied by the absolute number of school-age children. This is accomplished by filtering population data, only displaying individuals between 5-19 years of age. This is shown in the formula below. The result is the projected number of illiterate school-age individuals for a given region.

\begin{equation}
\begin{aligned}
\textrm{VbS 3: Education} = p_{i,school} * L_{I} * A_{i}
\\ 
\textrm{with} \hspace{0.5cm} L_{I} = \overline{L_{i}} \hspace{1cm} \forall i\in I
\label{normalize_edu}
\end{aligned}
\end{equation}
\vspace{0.2cm}

Lastly, the inverse of the USAID water access indicator $W$ provides data for VbS 4: Water demand.

\begin{equation}
\begin{aligned}
\textrm{VbS 4: Water} = p_{i} * W_{i}* A_{i} \hspace{1cm} \forall i \in I
\end{aligned}
\label{normalize_water}
\end{equation}


\subsection{Vehicle service stop allocation}
\label{sec:vehicle_stop_locations}
Vehicle service stop locations are derived exclusively using WorldPop regional population data. This results in constant vehicle service stops that retain validity regardless of the specific combination of services offered on a particular route, assuming a fixed population distribution. To locate vehicle stop locations, we utilize a k-means clustering algorithm. A cluster is characterized by a center, the cluster centroid, and the population assigned to this center. Compared to other clustering methods like hierarchical attribute-based clustering and density-based spatial clustering, this approach comes with several advantages: (1) Resulting clusters are computationally straightforward to group; (2) Continually good performance on large datasets; (3) Clusters are symmetric and spherically shaped allowing for centrally located vehicle stops. The k-means algorithm requires the user to input the desired number of clusters, which then correspond to the number of randomly initiated centroids; the algorithm subsequently groups the data by assigning each population pixel to its closest respective cluster centroid. As the algorithm progresses, the centroids of each evolving cluster are continuously updated, and the surrounding data is reassigned to the closest centroid; this process is repeated until a stable solution is found. 

\begin{figure}[htb]
\includegraphics[scale=0.5]{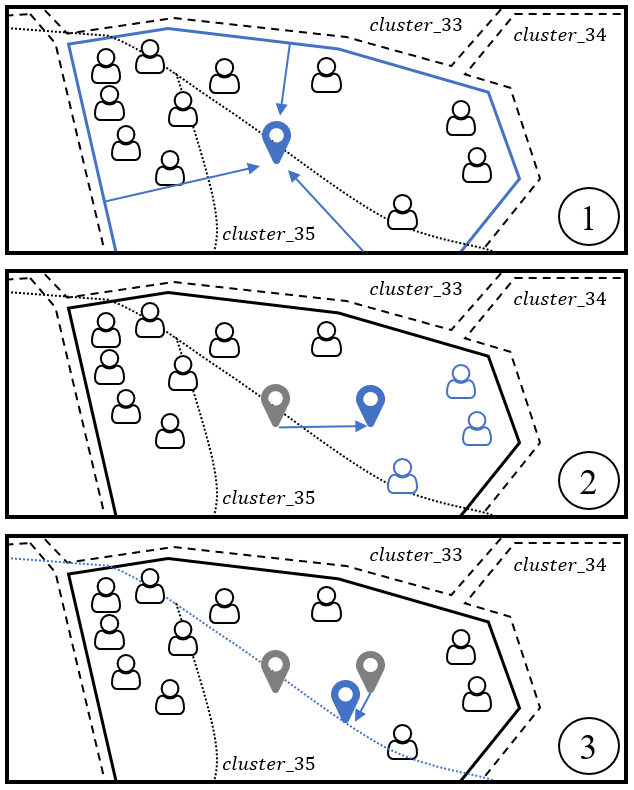}
\centering
\caption{Overview three step vehicle service stop allocation model}
\label{process3}
\end{figure}

Using the mean coordinate function in QGIS, we can calculate the respective cluster center of mass for every iteration of the k-means algorithm (step 1, Figure \ref{process3}). This is an essential component of evaluating clusters based on accessibility. In principle, each demand pixel is assigned to one cluster centroid before the mean distance of all demand pixel to their respective centroid is calculated. We further incorporate a demand attribute weighting factor to reduce individual travel distance to the closest service stop. Its effect is pulling the given mean coordinates towards areas of projected unmet demand, further from assumed service centers (step 2, Figure \ref{process3}). An increasing number of cluster centroids represents a trade-off between increasing operational costs and spatial accessibility. Because of this trade-off, the smallest set of stop locations that can satisfy the greatest need for a given service while minimizing the overlap between catchment areas needs to be determined. Following our initial assumptions, no cluster should have an average straight-line distance to the center point exceeding 5 km. We incrementally increase the total number by 10 to derive a fitting number of clusters. Once a final number of clusters is determined, we assess whether center points are on the existing road network (step 3, Figure \ref{process3}). 

Further, all potential vehicle service stop locations must reside inside the assumed daily vehicle range \cite{Pizzinini2023VehiclesService}. Outliers are manually removed, similar to the approach introduced by Straitiff et al. \cite{Straitiff2010UsingPlane}. More importantly, the inability to accurately project travel time and distances between vehicle service stops that do not lie on established transport routes would significantly hinder the ability to conduct subsequent routing optimizations and may lie on terrain that is not traversable. Using this set of adjusted vehicle service stops, we utilize the buffer tool in QGIS to calculate the absolute demand inside each location`s catchment area. Thereby, four service demand values per pixel inside the 5 km catchments are summed up. We employ Voronoi polygons to divide intersecting catchment areas to avoid double-counting populations living in the overlapping regions of two buffer zones. This ensures that each demand value is attributed only to its closest cluster center point. The result is the summed service demand per cluster, which can be displayed in an attribute table for every service under analysis. This grouping of service-specific demand data by vehicle stop location allows local authorities and fleet operators to effectively prioritize areas where the population is least likely to access public services, thus targeting the highest share of unmet demand. To derive a coverage value for each service, we divide the total covered service demand that is equal to the sum of all pixels inside the 5 km catchments by the total demand value that was calculated for all pixels in section \ref{demandquantification}. 

\section{Case study results}
\label{sec:results}

\subsection{Unmet and addressable demand}
\begin{figure}[htb]
\includegraphics[width=13cm]{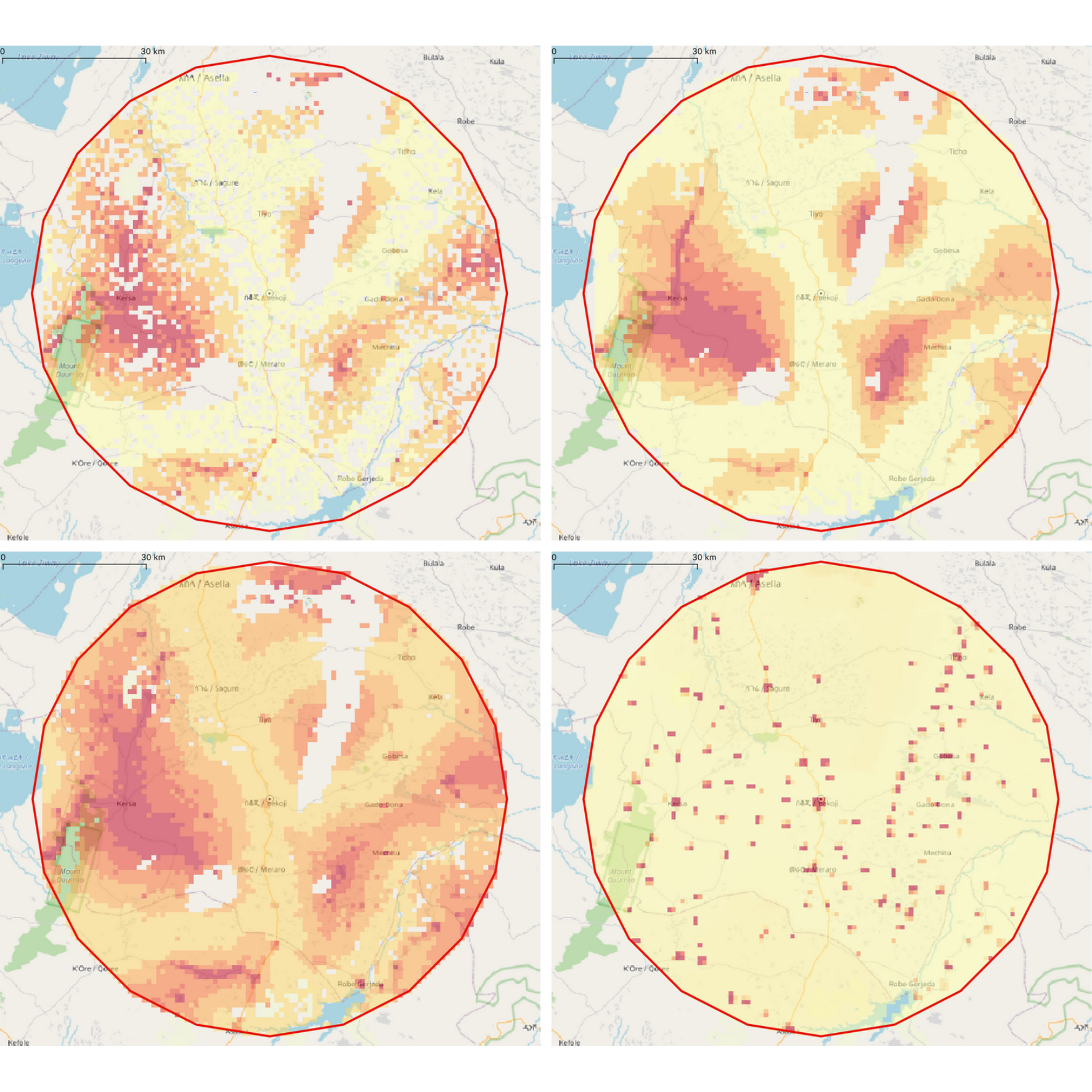}
\caption{Results for the demand analysis with red = high demand and yellow = low demand. Upper left: education, upper right: water, lower left: energy, lower right: ANC} 
\label{DemandOverview}
\end{figure}

The case study area yields a total population of 1,919,790 individuals. Applying our methodology in Bekoji results in 163,346 discrete raster cells. Each cell is coupled with demand data for the four introduced public services. Figure \ref{DemandOverview} illustrates the demand distribution of the four public services as a gradient from red (= high demand) to yellow (= low demand) with Bekoji at its center point. It is important to consider our demand quantification process again, where absolute demand for each service is multiplied by the Accessibility Scaling Factor. This generally balances results for high-demand and high-accessibility relative to low-demand and low-accessibility regions. For three of the depicted distributions, it is clear that service demand is low along the main north-south road connecting Bekoji to the capital Addis Ababa. Although the area around this road network is populated, communities along such infrastructure have relatively well-established access to water, education, and electricity supply. This holds for all buffer zones along the other sub-county roads connecting Bekoji to the west and east of the Oromia region. In between, demand for these services increases as accessibility decreases. Following our analysis, about 409,335 individuals lack access to electricity, 234,375 individuals experience insufficient access to clean drinking water, and 117,343 school-aged children lack the required education materials. Only demand from about 52,743 women for ANC is relatively homogeneously distributed across the target area (see table \ref{result}). This can be attributed to two factors. First, in absolute numbers, demand for ANC is lower than for the other three services. Second, primary healthcare facilities can be found in many of the remote communities \cite{Ouma2018AccessAnalysis}. This only renders a few scattered population hotspots inaccessible, as Figure \ref{DemandOverview} illustrates. 

Because vehicle service stop locations must be located on existing road infrastructure, we consider the population living further than 5 km from a traversal road inaccessible and excluded from further analysis. Our analysis shows that over 77\% of the total population lives within 1 kilometer of the nearest road, and 94\% within 5 km. We consider these demand points as adressable demand. Table \ref{result} gives an overview of the adressable demand per service. The high share of population with proximity to traversal roads is a first indication that VbS can be a viable intervention for public service delivery. 


\subsection{Vehicle stop locations}
\subsubsection{Cluster definition}
\begin{figure}[htb]
\centering
\includegraphics[width=9cm]{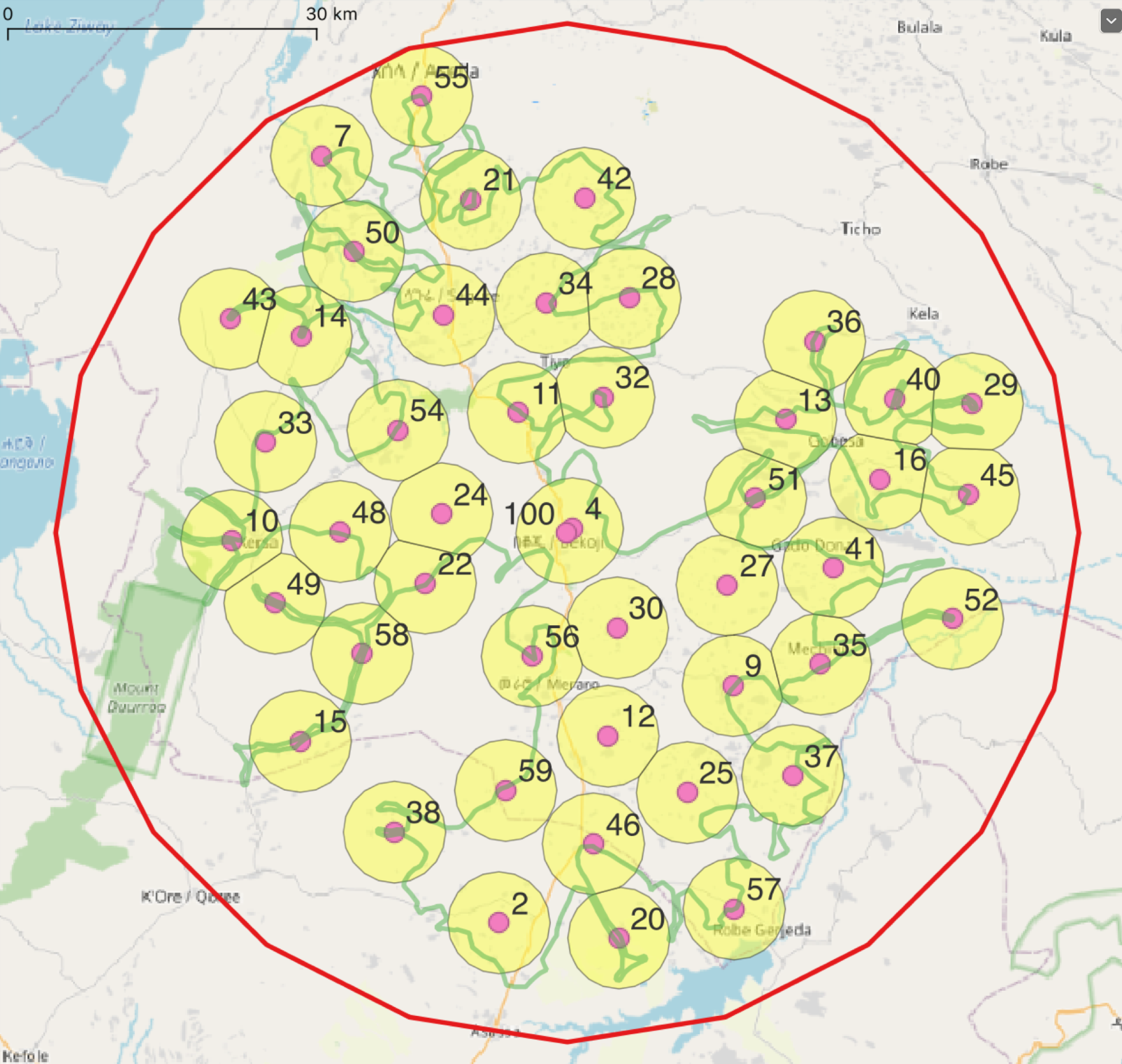}
\caption{The resulting 45 final vehicle service stops (= red points with yellow 5 km catchment area) for the case study in Bekoji}
\label{finalvehiclestops}
\end{figure}

After undergoing the requisite iterations to ensure that the maximum average cluster distance remained below 5 km, we computed centroid locations following the procedures outlined. The resulting vehicle service stops, are depicted in Figure \ref{finalvehiclestops}. ClusterID 100 designates the origin point at Bekoji. Coordinates for all locations are provided in the Appendix.

Two factors influence the absolute count and distribution of destinations. Firstly, the population distribution determines the number of iterations necessary to meet the minimum cluster distance. As in Ethiopia, regions with a more even population distribution would inherently require fewer clusters. Secondly, the service range, contingent on factors such as the vehicle model, local road networks, and regional terrain, dictates the realistic accessibility of derived destinations within the area of interest. Considering these boundary conditions, the final destination sets for Bekoji were curtailed to 45 vehicle service stops with an average straight-line travel distance to the cluster centroid of 3.97 km, well below the defined maximum of 5 km. 

\begin{figure}[htb]
\includegraphics[width=12cm]{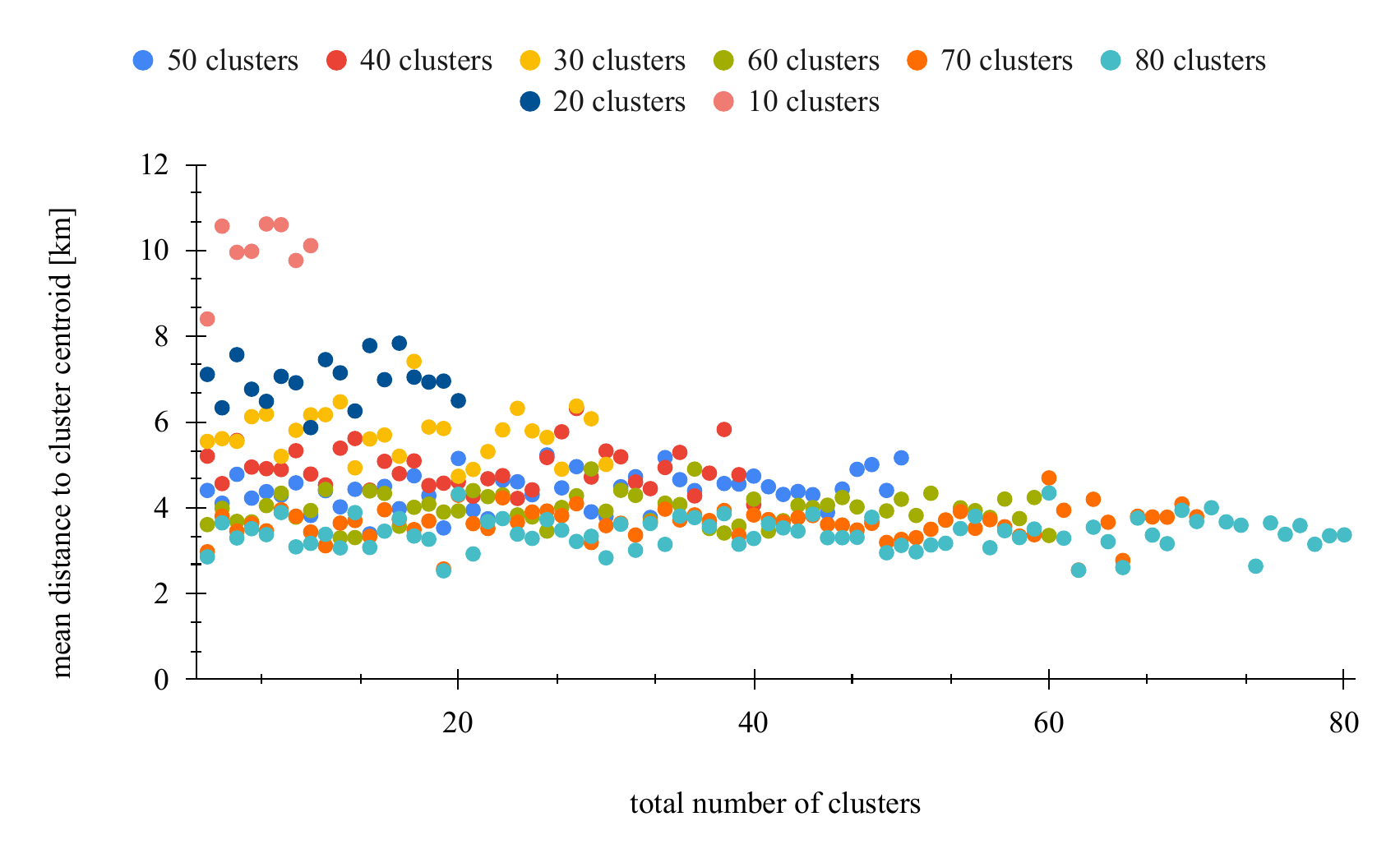}
\caption{Overview of mean distance to cluster center of different cluster sets} 
\label{entwicklung}
\end{figure}

More interestingly, this case study demonstrates an apparent diminishing returns phenomenon over the number of vehicle service stop locations (Figure \ref{entwicklung}). As the number of clusters increases, the additional service coverage attained per additional vehicle service stop decreases. With iteration increments of 10 clusters, 60 clusters represent the minimal number of locations to satisfy a 5 km maximum travel distance threshold. This effect is inherent to the applied k-means algorithm. The clustering algorithm aims to minimize the sum of squared distances between population points and their assigned cluster centroids. As the number of clusters increases, the distance decrease. However, beyond a certain point, the reduction becomes marginal. 

It is crucial to emphasize that this diminishing returns phenomenon has practical implications for optimizing service coverage. The findings suggest that there exists an optimal balance between the number of vehicle service stop locations and the associated achieved accessibility. Beyond this optimal point, the marginal benefits of introducing additional stops diminish, potentially highlighting the need for a more nuanced approach to the placement of vehicle service stops in order to further optimize the trade-off between coverage and efficiency. 

\begin{figure}[htb]
\includegraphics[width=12cm]{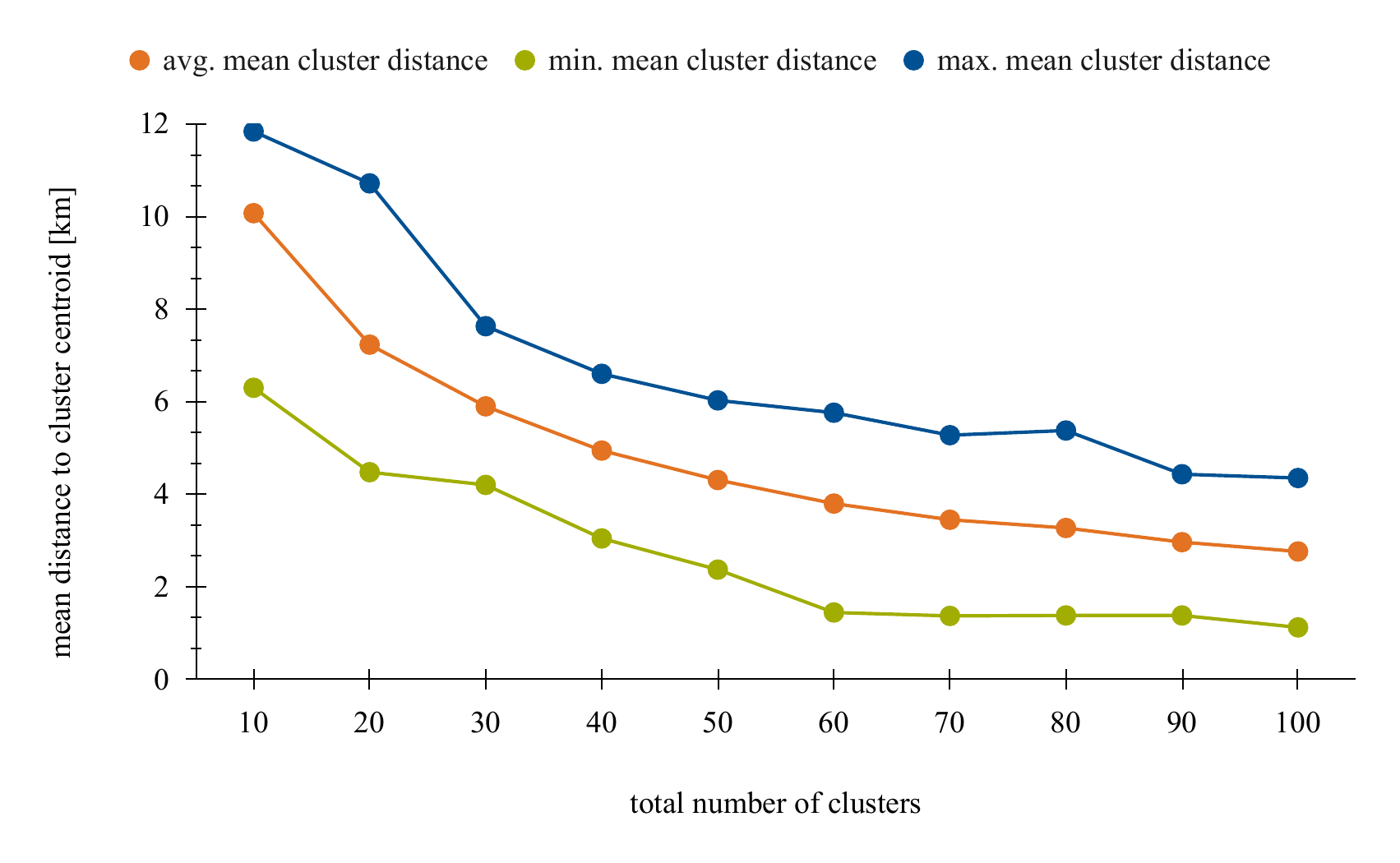}
\caption{Mean cluster distance decreases with increasing amount of clusters.} 
\label{accumuliert1}
\end{figure}

Critical indicators for our study's outcomes are the average, minimal, and maximal mean cluster distances, as illustrated in Figure \ref{accumuliert1}. As the number of clusters increases incrementally from 10 to 20 a substantial reduction in the average travel distance is noted. This reduction, however, exhibits a diminishing effect with successive iterations.

Conventionally, the optimal number of clusters in a k-means algorithm tends to converge towards demand hotspots, aligning clusters with prominent population centers. However, in the context of a relatively evenly distributed demand structure within the target area, the algorithm deviates from this norm. Instead of converging towards an optimal k that reflects inherent demand structures, it exhaustively disperses vehicle stop locations across the entire geographical expanse.

This departure from the conventional clustering behavior highlights the influence of the demand distribution on the algorithm's output. The drastic reduction in average travel distance suggests an initial clustering response to evenly distributed demand, emphasizing the algorithm's adaptability to varied demand scenarios. However, as iterations progress, the diminishing returns observed in the average travel distance underscore the need for a nuanced understanding of the interplay between demand distribution and clustering outcomes.

\subsubsection{Postprocessing}
In the last step, we evaluated whether the initial set of vehicle service stops is located on the road network (step 3, Figure \ref{process3}). If not, the centroid is put manually on the next road. The distance of this adjustment is, thus, at least in part, a function of local road network coverage, the regional population distribution, and the selected clustering algorithm. The median of these adjustments in Bekoji amounts to 571 m. While these modifications will slightly impact the coverage projections when compared to the original destinations, this discrepancy was accounted for by only summing the actual demand around the final destination set after the adjustment had been carried out. Future work on additional methods to minimize this would be valuable. 

During the computation of the centroids, a demand attribute weighting factor was introduced. This factor guides a given mean coordinate towards regions with anticipated unmet demand. We conducted a comparative analysis, considering scenarios with and without the weighting factor. It revealed that the inclusion of this factor resulted in a further 2.14\% reduction in the average Euclidean distance to the nearest centroid. Incorporating a demand-based weighting factor thus contributes to a more refined clustering outcome.

\subsubsection{Demand coverage}

\begin{table}[htb]
\centering
\caption{Results of the demand analysis performed for Bekoji. Total demand = total number of service customers within a 50 km radius around Bekoji; addressable demand = fraction of the total demand within 5 km of a traversal road; covered demand = demand within a 5 km Euclidean catchment of a vehicle service stop}
\setlength{\tabcolsep}{10pt} 
\renewcommand{\arraystretch}{1.5} 
\settowidth\rotheadsize{\theadfont Dynamic Read} 
\begin{tabularx}{\textwidth}{ ccccccc }
\hline
\thead{VbS} &  
\rothead{input data} &  
\rothead{accessibility measure} &
\rothead{total demand} &
\rothead{addressable demand} &
\rothead{covered demand} &
\rothead{covered demand} \\ \hline
ANC & \cite{OSM2022OpenStreetMaps, Linard2012Population2010, Tatem2014MappingBirths} & direct & 52,743 & 52,060 & 20,536 & 38.93\% \\
energy & \cite{OSM2022OpenStreetMaps, Linard2012Population2010, NASAEarthObservatory2022EarthMaps} & proxy & 409,335 & 404,548 & 239,160 & 58.42\% \\
education & \cite{OSM2022OpenStreetMaps, Linard2012Population2010, Tatem2014MappingBirths, USAID2022TheRepository} & proxy & 117,348 & 115,801 & 71,064 & 60.55\% \\
water & \cite{OSM2022OpenStreetMaps, Linard2012Population2010, Tatem2014MappingBirths, USAID2022TheRepository} & proxy & 234,375 & 231,159 & 152.095 & 62.49\%\\
\end{tabularx}
\label{result}
\end{table}

\begin{figure}[htb]
\includegraphics[width=12cm]{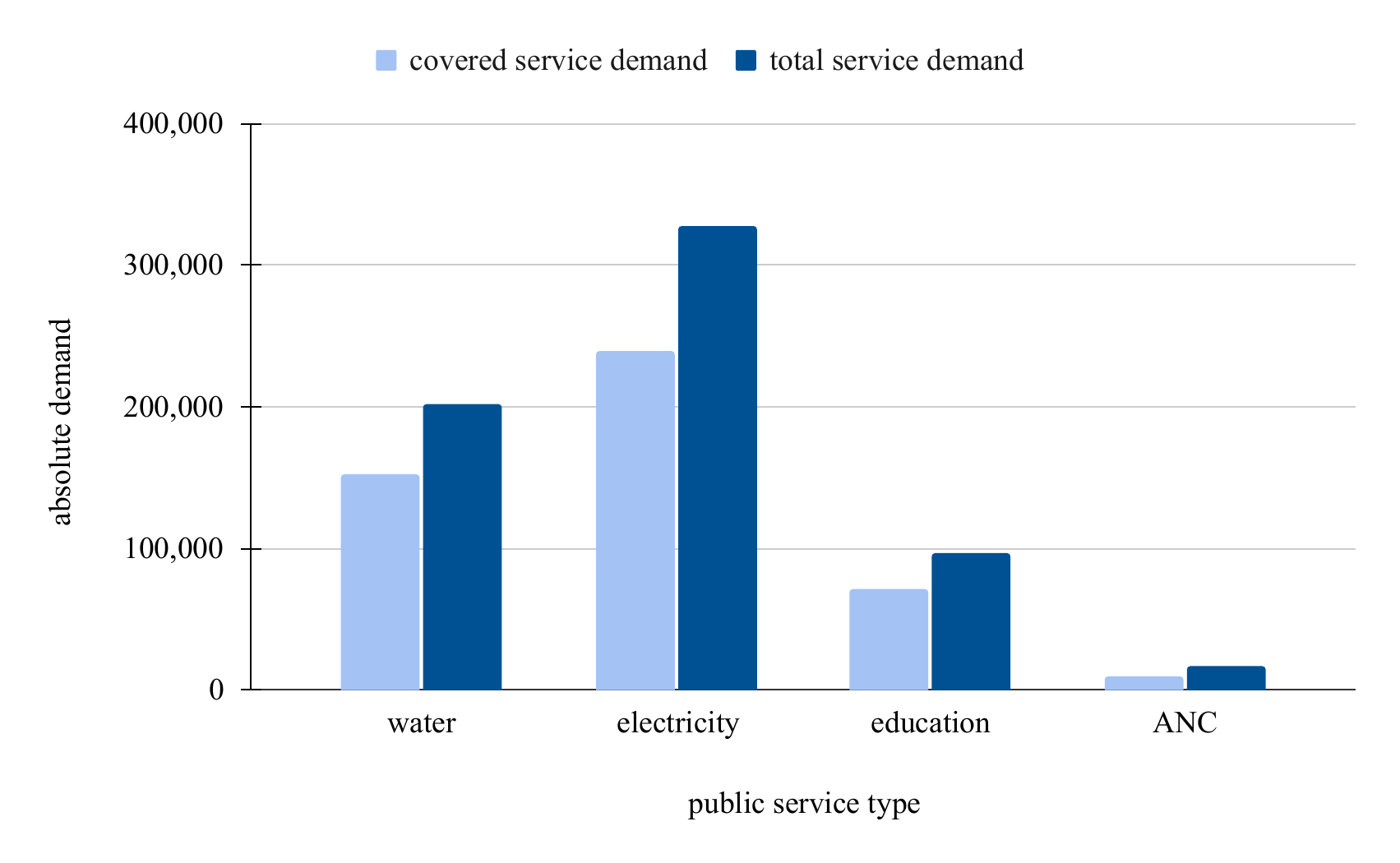}
\caption{Total and covered demand per public service type} 
\label{demandcoveragebalken}
\end{figure}

Out of this subset of addressable demand, the located vehicle service stops cover varying demand for each service (Figure \ref{demandcoveragebalken} and Table \ref{result}). The average vehicle service stop with its 5 km catchment area and four service offerings serve 10,496 demand points which comprise of 27\% water, 50\% energy, 15\% education, and 8\% ANC. It is important here to notice, that one individual can represent up to four demand points. 

Further, demand points for each service have different demand frequencies. Whereas the supply of 5 l drinking water reoccurs daily, ANC checkups are recommended three times during a nine month pregnancy \cite{WorldHealthOrganization2003ManagingDoctors}. Operational considerations about how often and at what time a vehicle service stop needs to be visited by the mobile facility are subject of the VbS fleet design. 

\subsubsection{Priorization}
\begin{figure}[htb]
\includegraphics[width=12cm]{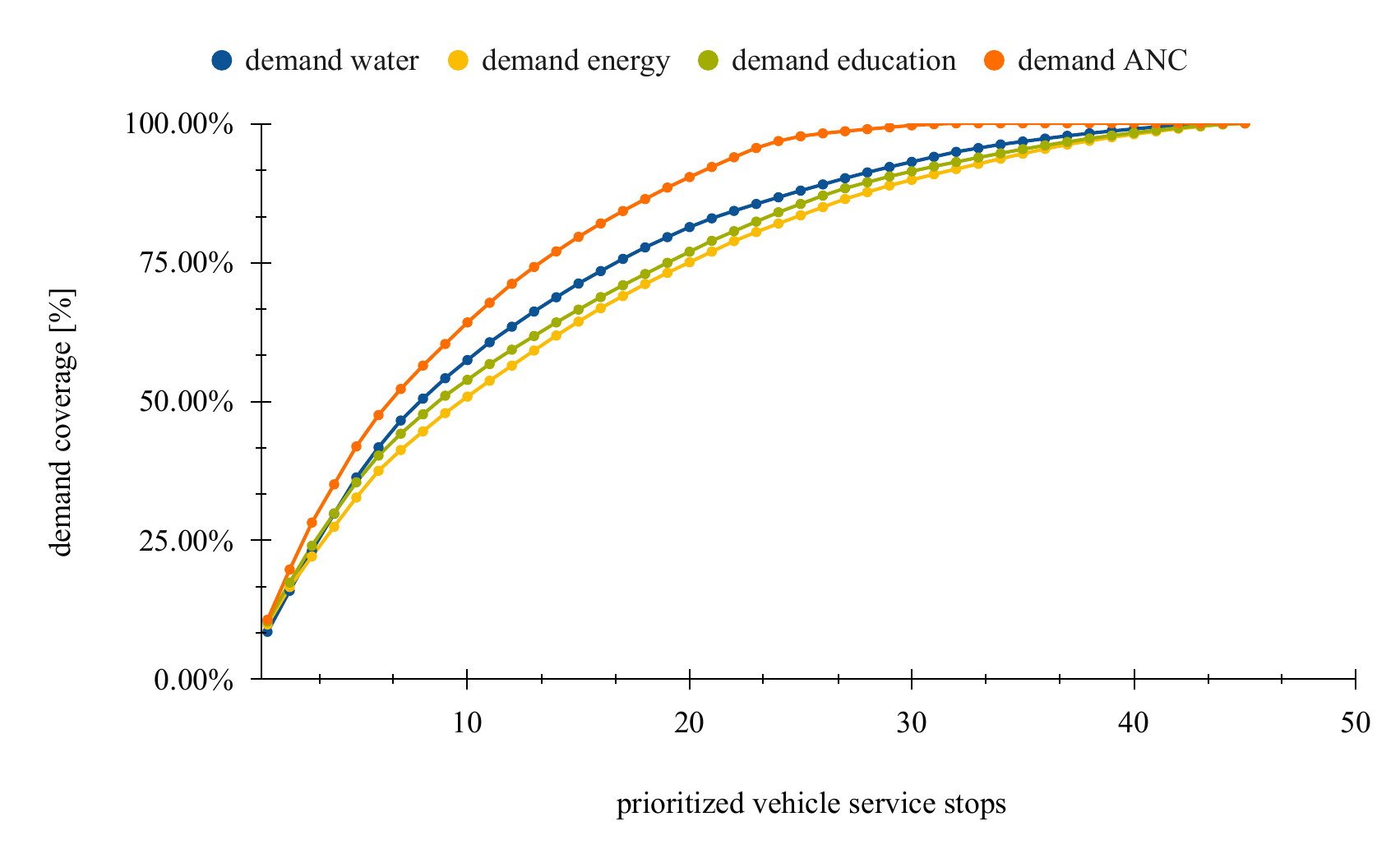}
\caption{Accumulated relative demand coverage over prioritized vehicle stops for the region around Bekoji with 100\% referring to the addressable demand} 
\label{accumuliert}
\end{figure}

Deriving the final vehicle service stop location set is an important step, however, under the assumption of limited resources, it will likely be the case that not all of these locations can be visited with the same regularity, if at all. This could result from any number of reasons; it is for example conceivable that limited funding only allows for a certain number of vehicles, which in turn limits the capacity to provide a greater quantity of a given service. Or it could be that personnel restrictions restrict the number of possible locations due to staffing shortages. In any case, it becomes apparent that a way to prioritize these vehicle stop locations is needed. It follows that of the total number of destinations required to satisfy the 5 km threshold, those with the greatest projected number of inaccessible individuals in need of a given service should be visited first. By summing the estimated demand for all services in all catchments it becomes possible to rank destinations according to the specific service coverage. This is illustrated in Figure \ref{accumuliert}. While a similar trajectory can be observed for all services, it is interesting to note that there are variations.  These variations are a function of regional population distribution and are a manifestation of the tailored vehicle service stop that make this approach applicable to rural environments with sparse information. In the case of ANC for instance, it becomes apparent that from the 32nd destination onwards, no additional gains in coverage are realized. Alternatively, it can be observed that a greater projected demand for drinking water is satisfied with fewer destinations compared to other services.

When it comes to determining the prioritization of destinations, again Figure \ref{accumuliert} is key. In general, two different ways present themselves along which to interpret these results. (1) From the x-axis; Assuming that the project has enough resources for a limited number of destinations, this plot yields a corresponding maximum achievable service demand coverage. The figure reveals which destinations should make up this limited number, and in what order they should be prioritized. (2) From the y-axis; A desired service demand coverage is stipulated in line with project objectives; in this case, this figure yields the number and the ID of the destinations which need to be visited to achieve this level of coverage most efficiently. To clarify then, while it is true that the stop locations are drawn from the same total pool of possible destinations, determined based on the region’s population distribution, the exact IDs, as well as the order in which these destinations are visited, will vary by the specific service under consideration. 

\subsubsection{OD matrice}
To create an actionable output, we apply the matrix-from-layers function available through the Open Route Service plugin in QGIS. Open Route Service is a crowdsourced, open-source planning tool that provides the possibility of calculating time or distance matrices between many different predefined destinations, so-called origin-destination matrices. In such a matrix, the travel times and real distances between all point combinations are computed (Table \ref{ODM}). The matrix starts at ID=0, and first computes all times and distances from this ID to all other set IDs, before moving on to ID=1 and again computing the time and distance to all other set IDs. This results in 2,116 different destination-to-destination combinations for our case study in the area around Bekoji.

\begin{table}[htb]
\centering
\caption{Example of the resulting OD matrice}
\begin{tabular}{P{1.5cm}P{1.5cm}p{1.5cm}p{2cm}p{2cm}}
\hline
 &  from ID & to ID & travel time (h) & distance (km) \\ 
\hline
1 & 0 & 0 & 0 & 0 \\
2 & 0 & 4 & 1.059 & 54.558 \\
3 & 0 & 5 & 0.767 & 29.260 \\
4 & 0 & 8 & 1.606 & 99.493 \\
5 & 0 & 9 & 0.846 & 32.836 \\
... & ... & ... & ... & ... \\
3481 & 100 & 89 & 0.388 & 13.114 \\
\end{tabular}
\label{ODM}
\end{table}

\subsection{Analytical Validation}\label{sec:analytical_validation}

\begin{figure}[htb]
\centering
\includegraphics[width=\textwidth]{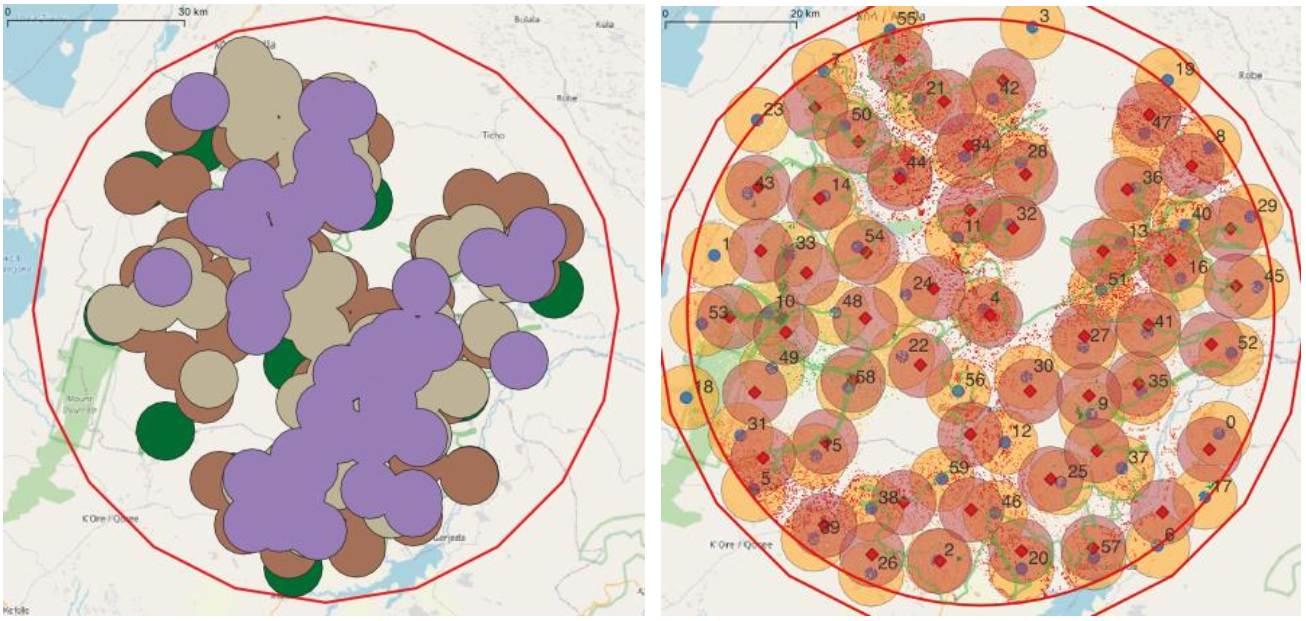}
\caption{Left: Results for randomized distributed vehicle service stops. Right: Reducing area of investigation}
\label{validationofresults}
\end{figure}

Any work attempting to model complex sociodemographic interactions using simplifying assumptions inherently risks misrepresenting local situations. The question becomes whether the simplifications made in this context prove useful in pursuit of the overarching goal of enhancing rural accessibility. We perform two evaluation approaches. First, we compare the resulting total coverage of 73\% to results stemming from randomized vehicle stop locations. Under four varying conditions, the same number of vehicle stop locations (= 45) is located, and the population residing within the 5 km catchment areas is summed. The results for the total coverage (\% of total demand) of randomized distributed stops are listed here: (1) completely random distribution (41.25\%); (2) distance between service stops $\geq$ 5 km (58.69\%); (3) distance between service stops $\geq$ 7.5 km (63.47\%); (4) distance between service stops $\geq$ 10 km (42.94\%); (5) k-means (72.84\%). Our approach covers 9\% more demand than any random selection with the same number of destinations, demonstrating the validity of this targeted approach (see Figure \ref{validationofresults}, left).

Secondly, we assess the result's sensitivity by adjusting the area of interest. While 100-250 iterations performed by k-means naturally converge to the same set of clusters - holding all parameters constant - this is largely attributable to the fact that the same population distribution was used for each run. Therefore, we decreased the 50 km Euclidean radius around Bekoji to 45 km, reducing the total area by 19\% and subsequently changing the total population distribution in our area of consideration (see Figure \ref{validationofresults}, right). Despite this reduction, leading to 10 fewer clusters achieving the same individual level of accessibility as before, an absolute population coverage difference of just 2\% is recorded. This is not to say that the same population covered previously is also covered by this new destination set; instead, it says that of the total population encompassed in this region of interest, a difference of less than 2\% in the overall number of people who live within 5 km of a cluster centroid was observed. Thus, while exact locations will inevitably vary depending on how a point set is cropped to a relevant area of interest, it appears that the resulting service coverage is comparatively insensitive to these changes and that this approach is robust to boundary variations. 

\subsection{Agenda for Action}\label{sec:agenda_for_action}
To support authorities and decision-makers during VbS project appraisal, we introduce some additional perspectives for consideration. The United Nations views geospatial data as a crucial component to guide Sustainable Development Goals related interventions, and it is estimated that approximately 20\% of all Sustainable Development Goals can be measured, directly or indirectly, through geospatial data \cite{Nations2019TheThemes}. Nevertheless, the potential of utilizing a data-driven location-allocation models may not always be apparent to regional authorities. The approach introduced in this work, in connection with the implementation of VbS, proves that regional accessibility challenges can be overcome in part through the effective utilization of such data. 

Plots such as Figure \ref{accumuliert} showing the effectivity-to-resource trade off as it pertains to the number of  vehicle stop locations, can serve as valuable tools for decision-makers. Some, for instance, may deem a level of service coverage lower than 100\% as regionally acceptable and dedicate remaining resources to other interventions such as upgrading the quality of service provided at given destinations \cite{Church1974TheProblem}. In any case, the utilization of geospatial information as it is outlined here makes these tradeoffs explicit and removes some of the uncertainty frequently associated with service allocation decisions.

Lastly, the underlying ethical allocation principle can be adapted to local perceptions of spatial justice. Whereas we combine an Egalitarianism  perspective (equal access) with a threshold service accessibility value (5 km accepted walking distance), authorities can adjust these assumptions within our methodological framework according to different interpretations of equality and spatial justice.

\section{Discussion \& Conclusion}\label{sec:conclusion}
Spatial accessibility - the ease with which individuals can move between different places - serves as a common denominator for many daily activities. In many cases, it is deceivingly simple to focus narrowly on a single intervention, when a lack of basic accessibility can cause even projects with the best of intentions to fail. From affordability and availability to acceptance and accommodation, accessibility has many dimensions; this work by no means does justice to all of them \cite{Guagliardo2004SpatialChallenges}. Instead, it builds on a premise that VbS harbors significant potential in bridging the accessibility gap which exists in so many rural communities, and looks to build on this framework. Specifically, this paper proposes a methodology that optimizes vehicle stop locations based on geospatial data for a variety of services and demonstrates the merits of this approach through a case study in Ethiopia. In contrast to other approaches, this work explicitly focuses on rural communities, where a lack of applicable solutions to addressing spatial access problems perpetuates a cycle of isolation. In contributing to the body of knowledge on this matter, this work suggests an alternative approach that has the potential to address the unique needs of these vulnerable populations more constructively.

The most notable limitation in existing literature was the requirement to predetermine a number of facility destinations; this runs the risk of perpetuating the very rural accessibility problems this work sets out to address. The clustering approach used here does not require this. Instead, it leverages the increasing functionality of GIS software to derive service-specific demand maps which build on existing regional population distributions and can incorporate primary data from a wide variety of sources; this allows for hundreds of thousands of granular population locations to flow into an analysis. Not only does this approach not require the selection of specific candidate destinations, but it also provides decision-makers with valuable information which visualizes this coverage data and makes explicit the trade-off between invested resources and regional service coverage. This lays the groundwork for more informed decisions that are tailored to regional characteristics and automatically account for factors such as local population distribution and existing service coverage. Furthermore, this research specifically addresses the problem of rural isolation; by scaling projected service demand with travel times, in line with the Rawlsian principle of spatial justice, the project thereby ensures that the approaches outlined here benefit the most vulnerable populations of a given region.

Building on the framework presented here, there are several promising directions through which future research could add to this work. One way would be to incorporate mobile phones more effectively, a technology that is widely utilized even in some of the most remote regions of the planet \cite{Oluwatobi2015MobilePoor, Wesolowski2015QuantifyingData}. Current projections of service demand in this work are inherently static and operate on the basis of predictable vehicle schedules. Finding a way of meaningfully incorporating time-sensitive, demand-specific information through mobile alerts, notifications, or similar mechanisms, could thereby more effectively address urgent needs otherwise left unattended; in addition, such features give community stakeholders a louder voice which may increase overall acceptance. Some programs incorporating such functionalities are already being implemented \cite{Oluwatobi2015MobilePoor}. By providing access to more granular population-specific information, this technology opens the possibility of further tailoring offerings to local needs, something which is highly relevant for the delivery of various services. Conducting surveys is notoriously time-consuming and expensive, while mobile phones yield significant information with little effort. For example, data may show that the majority of a given target population spends their day in a field far outside their village. This would mean that the assumed population distribution for a region is shifted during these times; taking this into account when deriving vehicle stop locations may further increase accessibility. It may well be the case that the ultimate challenge here lies not in accessing this data, but in dealing with, and parsing through, an information overload that may follow its implementation \cite{Sargent2010VerificationModels}.

In this research, various services such as drinking water and electricity are addressed independently of each other; however, using the same population base layer to estimate service demand, means that the final destination sets are drawn from an identical pool of potential stops for a given region. This opens the door for combinations of compatible services to be delivered simultaneously. It may be possible, for instance, to offer a service that provides ANC while also administering vaccinations. This would not only reduce costs but increase the reach of life-saving interventions. Exploring such possible combinations in greater detail through additional research would be valuable. 

When it comes to addressing some assumptions made in this work, there is promising potential to more accurately model the travel dynamics to various vehicle stops. While incorporating any generalizable insights about multi-modal transportation to these hubs would likely prove quite difficult, replacing an overarching 5 km radius assumption with a time or distance-based isochrone may further improve accuracy. Specifically, such an approach could more accurately account for local geography, thereby yielding more realistic service coverage maps, and subsequently influencing clustering iterations. The central assumption that our vehicle must always return to a central hub to replenish whatever service it has distributed, could also be examined through further research. It is possible that some services such as water or healthcare, for instance, do not require the vehicle to return to its start location, as multiple points exist at which a service can be refilled. This would in turn affect the assumed vehicle range as well as routing optimizations and may show that a single vehicle can access a wider area than previously assumed.

The problem of addressing the rural accessibility gap is a complex one. The research presented here, along with the improvements outlined above, are a small part of a continuously expanding body of knowledge, and with a constantly moving target, it is unlikely that research will ever converge on one optimal solution. It is the pursuit of this optimum that matters, however. From infant mortality to education, from economic well-being to food security, lowering the transaction costs of accessibility is the bedrock of many other interventions, and its ripple effects extend far beyond any single service or community \cite{Fouracre2007ACountries}. In this spirit, it is important to continue to facilitate this iterative learning. By coupling the implementation of a methodology like the one outlined here to objective impact assessments that evaluate its performance and continue to challenge its assumptions, it is possible to collectively ensure that the work being done in this space continues to improve the lives of those that need it most.

\backmatter

\bmhead{Acknowledgments}

First author C. P. devised the idea for this paper and drafted the research proposal. The second author N. J. conducted his research under the supervision of C. P. and based on the given research proposal. C. P. conducted the literature review and drafted the introduction. The methodology, case study, and discussion sections are based on the findings of N. J. and reworked by C.P. The paper's fieldwork was accompanied by D.Z., who also reviewed and formatted the paper. M. L. made an essential contribution to the conception of the research proposal. He critically revised the paper for its important intellectual content. M.L. gave final approval of the version to be published and agrees to all aspects of the work. All authors have read and agreed to the published version of the manuscript. 

\section*{Declarations}
This research was accomplished through and funded by the German Federal Ministry for Economic Cooperation and Development (BMZ). The authors declare no conflict of interest between funding and the presented research approach.

\begin{appendices}
\section{Data Preparation}\label{appendix1}

We encourage open source and make all relevant data for replication of the introduced approach available online.

\begin{table}[htb]
\caption{\label{tab:qgis_sql}QGIS SQL requests for service-specific demand retrieval}
\setlength{\tabcolsep}{10pt} 
\renewcommand{\arraystretch}{1.5} 
\renewcommand\tabularxcolumn[1]{m{#1}}
\begin{tabularx}{\textwidth}{ clX}
\hline
Nr. & VbS & SQL Command \\ \hline
1 & ANC & n/a \\
2 & energy & [No. of individuals per pixel * (night light data intensity per pixel / maximum registered light intensity in the area of interest)] * (travel time to the nearest large settlement per pixel (min) / maximum travel time in the area of interest (min)) \\
3 & education & [No. of school-age children (5-19 yrs) per pixel * (1 - avg. literacy rate (\%) per pixel) +No.of women of childbearing age (15-49 yrs) per pixel * unmet need for family planning (\%) per pixel] * (travel time to the nearest large settlement per pixel (min) / maximum travel time in the area of interest (min)) \\
4 & water & 1 - access to an improved water source \% per pixel* No. of individuals per pixel * (travel time to the nearest large settlement per pixel (min) / maximum travel time in the area of interest (min)) \\
\end{tabularx}
\end{table}

\begin{table}[htb]
  \caption{\label{tab:table-name}Overview of vehicle service stops including location and demand quantities for Bekoji, Ethiopia}
  \label{ODM tabel}
  \centering
  \renewcommand{\arraystretch}{0.5}
    \begin{tabular}{|r|r|r|rrrr}
    \toprule
    \multicolumn{1}{|p{3.715em}|}{ClusterId} & \multicolumn{1}{p{5.785em}|}{Latitude} & \multicolumn{1}{p{5.785em}|}{Longitude} & \multicolumn{1}{p{5.785em}|}{Demand Water} & \multicolumn{1}{p{5.785em}|}{Demand energy} & \multicolumn{1}{p{5.785em}|}{Demand Education} & \multicolumn{1}{p{5.785em}|}{Demand ANC} \\
    \midrule
    2     & 719.357.615 & 3.919.800.918 & \multicolumn{1}{r|}{2.806.157.477} & \multicolumn{1}{r|}{684.100.195} & \multicolumn{1}{r|}{1.637.105.397} & \multicolumn{1}{r|}{4.037.666.044} \\
    \midrule
    4     & 7.542.189.312 & 392.626.937 & \multicolumn{1}{r|}{1.501.658.731} & \multicolumn{1}{r|}{2.221.451.024} & \multicolumn{1}{r|}{5.656.702.209} & \multicolumn{1}{r|}{3.392.385.783} \\
    \midrule
    7     & 7.871.920.597 & 3.904.199.972 & \multicolumn{1}{r|}{7.607.570.588} & \multicolumn{1}{r|}{366.222.699} & \multicolumn{1}{r|}{106.060.763} & \multicolumn{1}{r|}{1.107.927.395} \\
    \midrule
    9     & 740.329.651 & 3.940.438.618 & \multicolumn{1}{r|}{1.121.911.949} & \multicolumn{1}{r|}{1.256.986.931} & \multicolumn{1}{r|}{3.408.343.443} & \multicolumn{1}{r|}{3.449.471.743} \\
    \midrule
    10    & 7.531.603.651 & 3.896.287.785 & \multicolumn{1}{r|}{1.302.979.209} & \multicolumn{1}{r|}{2.365.583.034} & \multicolumn{1}{r|}{7.390.624.885} & \multicolumn{1}{r|}{4.267.857.499} \\
    \midrule
    11    & 7.644.799.018 & 3.921.524.847 & \multicolumn{1}{r|}{5.491.227.201} & \multicolumn{1}{r|}{9.570.692.997} & \multicolumn{1}{r|}{252.703.409} & \multicolumn{1}{r|}{5.840.723.317} \\
    \midrule
    12    & 7.358.613.872 & 3.929.409.679 & \multicolumn{1}{r|}{1.368.525.571} & \multicolumn{1}{r|}{2.223.905.649} & \multicolumn{1}{r|}{5.329.803.271} & \multicolumn{1}{r|}{5.447.677.252} \\
    \midrule
    13    & 7.638.917.643 & 394.503.673 & \multicolumn{1}{r|}{3.902.569.111} & \multicolumn{1}{r|}{5.809.323.272} & \multicolumn{1}{r|}{1.599.323.931} & \multicolumn{1}{r|}{853.928.566} \\
    \midrule
    14    & 771.242.491 & 3.902.382.663 & \multicolumn{1}{r|}{3.138.069.886} & \multicolumn{1}{r|}{7.852.956.087} & \multicolumn{1}{r|}{237.261.554} & \multicolumn{1}{r|}{482.532.547} \\
    \midrule
    15    & 7.354.071.508 & 3.902.307.439 & \multicolumn{1}{r|}{3.405.779.024} & \multicolumn{1}{r|}{5.154.554.459} & \multicolumn{1}{r|}{1.746.422.288} & \multicolumn{1}{r|}{6.768.498.843} \\
    \midrule
    16    & 7.585.656.721 & 3.953.341.402 & \multicolumn{1}{r|}{2.042.401.477} & \multicolumn{1}{r|}{4.525.902.236} & \multicolumn{1}{r|}{1.230.913.799} & \multicolumn{1}{r|}{1.523.025.413} \\
    \midrule
    20    & 718.012.535 & 3.930.369.663 & \multicolumn{1}{r|}{2.845.119.135} & \multicolumn{1}{r|}{9.383.972.105} & \multicolumn{1}{r|}{1.997.922.526} & \multicolumn{1}{r|}{5.761.372.023} \\
    \midrule
    21    & 7.833.223.867 & 3.917.304.322 & \multicolumn{1}{r|}{6.057.323.486} & \multicolumn{1}{r|}{1.584.239.852} & \multicolumn{1}{r|}{4.338.749.503} & \multicolumn{1}{r|}{1.268.921.306} \\
    \midrule
    22    & 7.494.035.284 & 3.913.308.587 & \multicolumn{1}{r|}{7.324.498.321} & \multicolumn{1}{r|}{8.008.003.186} & \multicolumn{1}{r|}{2.482.714.537} & \multicolumn{1}{r|}{6.012.996.078} \\
    \midrule
    24    & 755.535.878 & 3.914.770.253 & \multicolumn{1}{r|}{3.749.602.426} & \multicolumn{1}{r|}{4.580.504.314} & \multicolumn{1}{r|}{1.408.766.219} & \multicolumn{1}{r|}{6.372.848.544} \\
    \midrule
    25    & 730.864.236 & 3.936.398.278 & \multicolumn{1}{r|}{4.262.913.102} & \multicolumn{1}{r|}{6.495.105.964} & \multicolumn{1}{r|}{1.513.104.378} & \multicolumn{1}{r|}{1.191.780.672} \\
    \midrule
    27    & 7.492.393.522 & 3.939.913.513 & \multicolumn{1}{r|}{5.568.744.156} & \multicolumn{1}{r|}{6.493.052.869} & \multicolumn{1}{r|}{1.842.178.712} & \multicolumn{1}{r|}{3.579.568.887} \\
    \midrule
    28    & 7.746.275.263 & 3.931.328.325 & \multicolumn{1}{r|}{2.746.902.945} & \multicolumn{1}{r|}{3.528.353.258} & \multicolumn{1}{r|}{1.066.394.386} & \multicolumn{1}{r|}{4.529.773.936} \\
    \midrule
    29    & 7.652.760.097 & 3.961.469.598 & \multicolumn{1}{r|}{1.808.746.963} & \multicolumn{1}{r|}{3.479.295.913} & \multicolumn{1}{r|}{1.223.666.708} & \multicolumn{1}{r|}{2.480.473.206} \\
    \midrule
    30    & 7.454.407.261 & 3.930.231.665 & \multicolumn{1}{r|}{1.838.621.175} & \multicolumn{1}{r|}{2.405.647.029} & \multicolumn{1}{r|}{6.540.371.698} & \multicolumn{1}{r|}{743.424.474} \\
    \midrule
    32    & 7.658.427.453 & 392.902.543 & \multicolumn{1}{r|}{5.991.470.312} & \multicolumn{1}{r|}{6.453.787.241} & \multicolumn{1}{r|}{2.013.948.055} & \multicolumn{1}{r|}{4.714.262.786} \\
    \midrule
    33    & 7.619.070.631 & 3.899.265.643 & \multicolumn{1}{r|}{8.242.540.054} & \multicolumn{1}{r|}{1.607.291.665} & \multicolumn{1}{r|}{4.990.880.126} & \multicolumn{1}{r|}{2.725.706.654} \\
    \midrule
    34    & 7.741.532.054 & 3.923.955.076 & \multicolumn{1}{r|}{5.807.872.619} & \multicolumn{1}{r|}{1.160.196.411} & \multicolumn{1}{r|}{3.069.265.386} & \multicolumn{1}{r|}{5.973.722.534} \\
    \midrule
    35    & 7.422.414.024 & 3.948.108.577 & \multicolumn{1}{r|}{3.342.460.004} & \multicolumn{1}{r|}{5.216.584.738} & \multicolumn{1}{r|}{1.434.572.317} & \multicolumn{1}{r|}{2.878.248.606} \\
    \midrule
    36    & 770.783.907 & 3.947.583.151 & \multicolumn{1}{r|}{4.148.466.617} & \multicolumn{1}{r|}{5.973.483.052} & \multicolumn{1}{r|}{173.401.241} & \multicolumn{1}{r|}{1.127.982.719} \\
    \midrule
    37    & 7.323.888.987 & 3.945.679.268 & \multicolumn{1}{r|}{1.380.173.565} & \multicolumn{1}{r|}{3.531.742.045} & \multicolumn{1}{r|}{7.719.377.544} & \multicolumn{1}{r|}{3.519.675.949} \\
    \midrule
    38    & 7.273.688.439 & 3.910.606.907 & \multicolumn{1}{r|}{1.028.567.049} & \multicolumn{1}{r|}{2.280.175.833} & \multicolumn{1}{r|}{6.239.093.513} & \multicolumn{1}{r|}{3.517.398.181} \\
    \midrule
    40    & 7.656.705.237 & 3.954.640.115 & \multicolumn{1}{r|}{795.446.064} & \multicolumn{1}{r|}{2.138.907.309} & \multicolumn{1}{r|}{5.707.700.886} & \multicolumn{1}{r|}{6.475.863.698} \\
    \midrule
    41    & 7.507.574.238 & 3.949.246.547 & \multicolumn{1}{r|}{4.861.384.368} & \multicolumn{1}{r|}{7.078.213.058} & \multicolumn{1}{r|}{2.026.038.053} & \multicolumn{1}{r|}{7.674.514.227} \\
    \midrule
    42    & 7.834.341.919 & 3.927.338.236 & \multicolumn{1}{r|}{1.701.227.193} & \multicolumn{1}{r|}{2.967.424.742} & \multicolumn{1}{r|}{940.222.289} & \multicolumn{1}{r|}{257.568.809} \\
    \midrule
    43    & 7.727.638.487 & 389.612.126 & \multicolumn{1}{r|}{1.704.386.406} & \multicolumn{1}{r|}{4.551.086.465} & \multicolumn{1}{r|}{1.395.399.449} & \multicolumn{1}{r|}{3.654.071.345} \\
    \midrule
    44    & 7.731.269.846 & 3.914.955.448 & \multicolumn{1}{r|}{3.846.342.953} & \multicolumn{1}{r|}{1.294.350.953} & \multicolumn{1}{r|}{3.308.517.504} & \multicolumn{1}{r|}{1.274.380.684} \\
    \midrule
    45    & 7.572.672.427 & 3.961.153.132 & \multicolumn{1}{r|}{4.972.059.791} & \multicolumn{1}{r|}{8.930.473.123} & \multicolumn{1}{r|}{2.795.604.506} & \multicolumn{1}{r|}{1.945.092.669} \\
    \midrule
    46    & 7.263.373.898 & 3.928.117.376 & \multicolumn{1}{r|}{9.838.671.679} & \multicolumn{1}{r|}{2.394.041.036} & \multicolumn{1}{r|}{4.853.873.786} & \multicolumn{1}{r|}{3.626.810.284} \\
    \midrule
    48    & 7.539.673.136 & 3.905.824.918 & \multicolumn{1}{r|}{9.928.507.674} & \multicolumn{1}{r|}{127.444.781} & \multicolumn{1}{r|}{4.136.047.148} & \multicolumn{1}{r|}{1.272.771.472} \\
    \midrule
    49    & 7.476.322.742 & 3.900.092.321 & \multicolumn{1}{r|}{1.096.172.895} & \multicolumn{1}{r|}{1.316.552.972} & \multicolumn{1}{r|}{4.700.482.302} & \multicolumn{1}{r|}{5.573.847.828} \\
    \midrule
    50    & 7.787.526.954 & 3.907.013.146 & \multicolumn{1}{r|}{1.579.846.146} & \multicolumn{1}{r|}{4.796.461.151} & \multicolumn{1}{r|}{1.438.457.269} & \multicolumn{1}{r|}{3.061.558.647} \\
    \midrule
    51    & 7.569.104.895 & 3.942.397.284 & \multicolumn{1}{r|}{1.881.092.641} & \multicolumn{1}{r|}{284.962.643} & \multicolumn{1}{r|}{741.809.371} & \multicolumn{1}{r|}{3.583.143.539} \\
    \midrule
    52    & 74.624.653 & 3.959.739.957 & \multicolumn{1}{r|}{7.030.766.587} & \multicolumn{1}{r|}{1.224.239.054} & \multicolumn{1}{r|}{4.430.942.026} & \multicolumn{1}{r|}{2.250.652.648} \\
    \midrule
    54    & 7.629.308.606 & 3.910.934.985 & \multicolumn{1}{r|}{2.390.813.394} & \multicolumn{1}{r|}{3.897.283.948} & \multicolumn{1}{r|}{1.195.646.107} & \multicolumn{1}{r|}{5.536.028.574} \\
    \midrule
    55    & 7.924.779.741 & 3.913.005.088 & \multicolumn{1}{r|}{5.950.921.792} & \multicolumn{1}{r|}{4.401.178.772} & \multicolumn{1}{r|}{1.309.926.281} & \multicolumn{1}{r|}{4.494.864.321} \\
    \midrule
    56    & 7.429.214.553 & 3.922.761.869 & \multicolumn{1}{r|}{1.385.308.106} & \multicolumn{1}{r|}{2.098.836.118} & \multicolumn{1}{r|}{5.350.968.867} & \multicolumn{1}{r|}{1.260.800.812} \\
    \midrule
    57    & 720.572.303 & 3.940.510.518 & \multicolumn{1}{r|}{2.625.737.573} & \multicolumn{1}{r|}{1.848.879.168} & \multicolumn{1}{r|}{3.473.675.794} & \multicolumn{1}{r|}{2.358.834.633} \\
    \midrule
    58    & 7.431.651.542 & 3.907.762.908 & \multicolumn{1}{r|}{1.010.324.671} & \multicolumn{1}{r|}{1.152.486.248} & \multicolumn{1}{r|}{3.976.930.953} & \multicolumn{1}{r|}{5.229.565.833} \\
    \midrule
    59    & 7.310.637.264 & 3.920.400.159 & \multicolumn{1}{r|}{8.079.042.132} & \multicolumn{1}{r|}{1.539.794.875} & \multicolumn{1}{r|}{3.765.460.855} & \multicolumn{1}{r|}{3.695.632.083} \\
    \midrule
    100   & 7.538.002.472 & 392.580.316 &       &       &       &  \\
    \cmidrule{1-3}    
    \end{tabular}%
\end{table}%
\end{appendices}

\clearpage
\bibliography{references}
\bibliographystyle{ieeetr}

\end{document}